\documentclass[conference]{IEEEtran}
\IEEEoverridecommandlockouts
\usepackage{cite}
\usepackage{amsmath,amssymb,amsfonts}
\usepackage{algorithmic}
\usepackage{graphicx}
\usepackage{textcomp}
\usepackage{xcolor}
\usepackage{color,soul}
\def\BibTeX{{\rm B\kern-.05em{\sc i\kern-.025em b}\kern-.08em
    T\kern-.1667em\lower.7ex\hbox{E}\kern-.125emX}}
\begin{document}

\title{Real-Time RFI Mitigation for the Apertif Radio Transient System}

\author{
\IEEEauthorblockN{Alessio Sclocco\IEEEauthorrefmark{2}}
\thanks{\IEEEauthorrefmark{2} These authors contributed equally to this work.}
\IEEEauthorblockA{\textit{Netherlands eScience Center} \\
Amsterdam, the Netherlands \\
a.sclocco@esciencecenter.nl}
\and
\IEEEauthorblockN{Dany Vohl\IEEEauthorrefmark{2}}
\IEEEauthorblockA{\textit{ASTRON, Netherlands Institute}\\ 
\textit{for Radio Astronomy}\\
Dwingeloo, the Netherlands \\
vohl@astron.nl}
\and
\IEEEauthorblockN{Rob V. van Nieuwpoort}
\IEEEauthorblockA{\textit{Netherlands eScience Center and}\\ \textit{University of Amsterdam}\\
Amsterdam, the Netherlands \\
r.vannieuwpoort@esciencecenter.nl}
}

\maketitle

\begin{abstract}
Current and upcoming radio telescopes are being designed with increasing sensitivity to detect new and mysterious radio sources of astrophysical origin. While this increased sensitivity improves the likelihood of discoveries, it also makes these instruments more susceptible to the deleterious effects of Radio Frequency Interference (RFI). The challenge posed by RFI is exacerbated by the high data-rates achieved by modern radio telescopes, which require real-time processing to keep up with the data. Furthermore, the high data-rates do not allow for permanent storage of observations at high resolution. Offline RFI mitigation is therefore not possible anymore. The real-time requirement makes RFI mitigation even more challenging because, on one side, the techniques used for mitigation need to be fast and simple, and on the other side they also need to be robust enough to cope with just a partial view of the data.

The Apertif Radio Transient System (ARTS) is the real-time, time-domain, transient detection instrument of the Westerbork Synthesis Radio Telescope (WSRT), and it is a perfect example of this challenging scenario. This system processes 73~Gb of data per second, in real-time, searching for faint pulsars and Fast Radio Bursts. Despite the radio quiet zone around WSRT, the generation of RFI is becoming increasingly part of anthropic activities, especially in a densely populated environment like the Netherlands where the telescope is located. Furthermore, our sky is populated by a growing number of satellites for world-wide telecommunication. Hence, the ARTS pipeline requires state-of-the-art real-time RFI mitigation, even if it contains a deep learning classifier to reduce the number of false-positive detections.

Our solution to this challenge is RFIm, a high-performance, open-source, tuned, and extensible RFI mitigation library. The goal of this library is to provide users with RFI mitigation routines that are designed to run in real-time on many-core accelerators, such as Graphics Processing Units, and that can be highly-tuned to achieve code and performance portability to different hardware platforms and scientific use-cases. Results on ARTS show that we can achieve real-time RFI mitigation, with a minimal impact on the total execution time of the search pipeline, and considerably reduce the number of false-positives.
\end{abstract}

\section{Introduction}\label{sec:introduction}

For a little over a decade~\cite{Lorimer2007Sci...318..777L}, Fast Radio Bursts (FRBs) have represented the source of highly discussed open questions among astronomy academics. 
FRBs are millisecond-duration, highly energetic, dispersed radio pulses emitted from distant regions of space-time, far beyond the vicinity of the Milky Way (MW). Without considering propagation effects like scintillation and scattering, FRBs follow a relatively simple time-frequency structure described as being broadband and of narrow width~\cite{LorimerKramer2004hpa..book.....L}. 

Similarly to what is observed for single pulses from pulsars\footnote{Pulsars are neutron stars of a few kilometers in radius, with a few times the mass of our Sun, and stable rotation periods duration varying between seconds~\cite{Tan2018ApJ...866...54T} to milliseconds~\cite{Hessels2006Sci...311.1901H}.}, as an FRB travels through regions of ionized medium, a frequency-dependent time delay is imposed on its signal, where longer wavelengths of the propagating signal are slowed down more intensely than shorter ones. This delay is quantified by a dispersion measure (DM), in unit of parsec per cubic centimeter (pc/cm$^3$), proportional to the number of free electrons along the line of sight:

\begin{equation}
\textrm{DM} = \int^{d}_{0} n_e(l) \mathit{dl},   
\label{eq::dm}
\end{equation}

\noindent where $n_e$ is the electron number density, $l$ is a path length, and $d$ is the distance to the FRB~\cite{Petroff+2019A&ARv..27....4P}. While DM values for most known pulsars place them within the MW when compared to electron-density models~\cite{CordesLazio2002astro.ph..7156C, Yao2017ApJ...835...29Y}, the large DM values measured for FRBs suggest an extragalactic origin.

Among other international efforts, the observation program ALERT\footnote{ALERT stands for Apertif Legacy Exploration of the Radio Transient Sky.} searches for FRBs using the Westerbork Synthesis Radio Telescope (WSRT) located in the Netherlands. The receivers of the WSRT were recently upgraded with phased-array feeds called Apertif, which along with the Apertif Radio Transient System (ARTS~\cite{vanLeeuwen2014htu..conf...79V})---a hybrid machine of combined Field-Programmable Gate Arrays (FPGAs) and Graphics Processing Units (GPUs)---allow to form $\sim10^3$ beams on the sky with a high duty cycle. The resulting data stream is searched in real-time using the AMBER pipeline~\cite{Sclocco+2016A&C....14....1S} on the ARTS cluster, generating a list of single-pulse candidates for any detection having a Signal-to-Noise Ratio (SNR) greater than a specified threshold.

\begin{figure}
    \centering
    \includegraphics[width=0.4\textwidth]{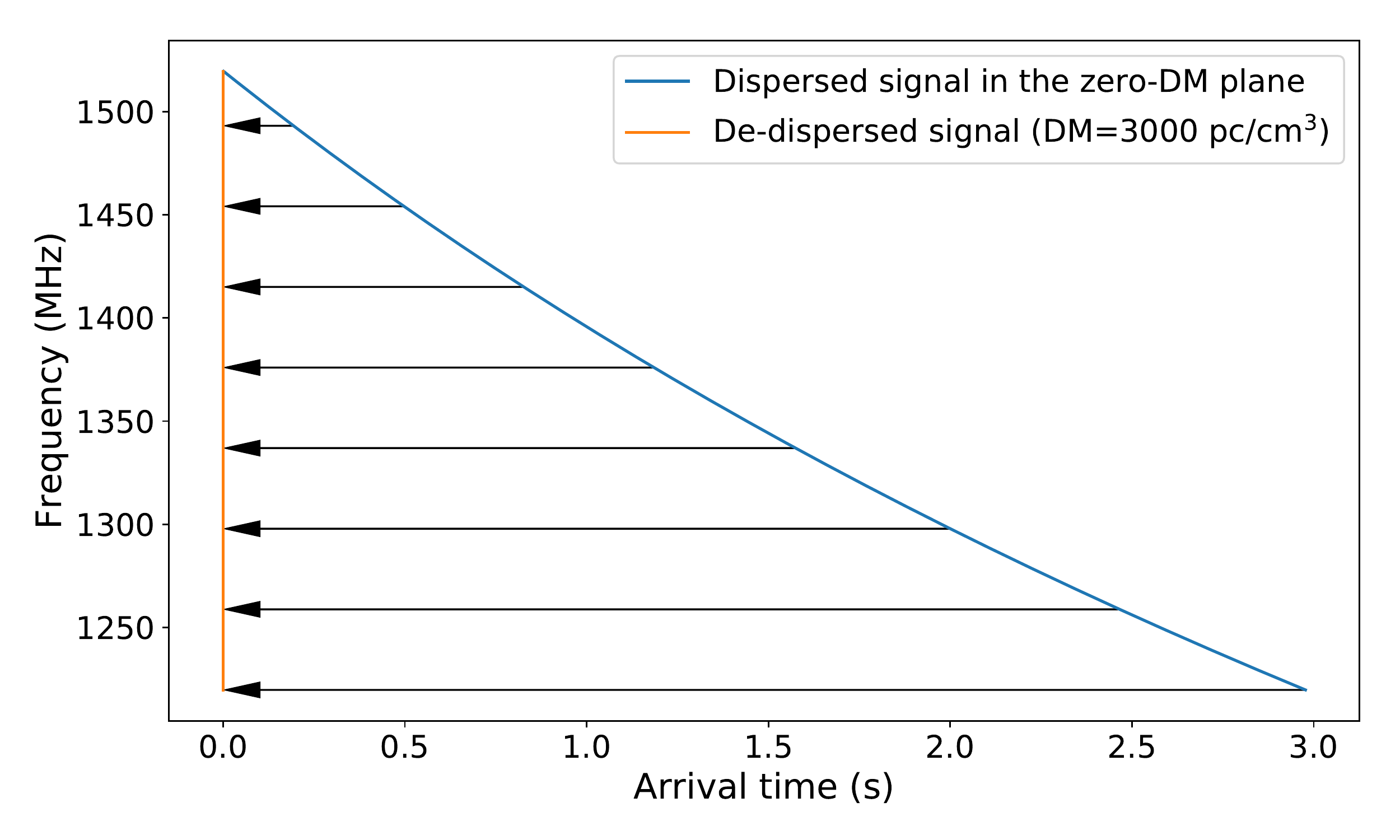}
    \caption{A signal dispersed in time as a function of frequency in zero-DM plane (blue), and the same pulse after incoherent de-dispersion (orange). The amounts of de-dispersion (Eq.~\ref{eq::delta_t}) applied to different frequency channels are shown by the black arrows.}
    \label{fig::dispersion}
\end{figure}

The common FRB searching strategy (such as~\cite{Barsdell2012MNRAS.422..379B, Sclocco+2016A&C....14....1S}) consists in incoherently de-dispersing the signal to correct for the time-delay among frequencies caused by dispersion (Figure~\ref{fig::dispersion}).
ARTS searches for FRBs by forming many different time series on the GPU corresponding to a wide range of dispersion measures.
The amount of dispersion in seconds is quantified by the time delay of the pulse between the lowest and highest radio frequencies of the observation in MHz, $v_{low}$ and $v_{high}$ respectively, as

\begin{equation}
\mathit{\Delta t} = k_{\textrm{DM}} \, (v_{low}^{-2}-v_{high}^{-2}) \, \textrm{DM} \;
\textrm{s},
\label{eq::delta_t}
\end{equation}

\noindent where $k_{\textrm{DM}}=\frac{e^2}{2\pi m_e c}\approx4.15\times10^3\, \textrm{MHz}^2\,pc^{-1}\,\textrm{cm}^3\,\textrm{s}$ is the dispersion constant, with $m_e$ being the mass of the electron, and $c$ the speed of light.

ARTS processes 73~Gb/s of Apertif data, in real-time, searching for faint pulsars and FRBs, while at the same time aiming to mitigate the effects of Radio Frequency Interference (RFI) in a densely populated environment like the Netherlands.
Hence, our real-time requirement translates to the need for RFI mitigation methods that are fast and simple, while also being robust enough to cope with a partial view of the data.

Even with a radio quiet zone around WSRT, the generation of RFI is becoming increasingly part of anthropic activities. Furthermore, our sky is populated by a growing number of satellites for world-wide telecommunications.
In fact, RFI can be stronger than astrophysical signals due to the inverse-square law of propagation which stipulates that electromagnetic radiation dissipates at a quadratic rate with respect to distance~\cite{boyle2019}. As opposed to the prototypical FRB signal, RFI is generated on or near Earth and displays a wide range of time and frequency structures not generally found in nature.

Without RFI mitigation, spurious bright individual RFI signals at different frequencies can artificially align at higher DMs, in which case the sum of these signals can become greater than the detection threshold, therefore causing artificial detections. 
These false positives can rapidly increase the size of the single-pulse candidates list, which needs to be verified by astronomers. 
Not only can RFI cause false-positive detections (i.e. non-astrophysical pulses erroneously classified as FRBs), but can also mask real, weak astrophysical signals and reduce the rate of true positives.
While a deep learning classifier~\cite{Connor+2018AJ....156..256C} is being used \emph{a posteriori} to reduce the number of false-positive detections that need to be verified by eye, removing RFI early on in the pipeline helps to reduce the overall compute time.

\section{Related work}
RFI mitigation has previously been tackled via different methods, including linear methods (e.g. Singular Value Decomposition~\cite{Offringa2010MNRAS.405..155O} and Principal Component Analysis~\cite{Li2006ITGRS..44..530L}), test of non-Gaussianity (e.g. Spectral Kurtosis~\cite{Nita2019JAI.....840008N, Taylor2019JAI.....840004T}), threshold-based methods (e.g. SumThreshold algorithm~\cite{offringa2012sumthreshold}, where RFI is defined as values above a defined threshold in the time-frequency plane), and more recently via machine learning methods like K-nearest neighbour, Gaussian mixture model, and convolutional neural network~\cite{Akeret2017A&C....18....8A, Akeret2017A&C....18...35A, Czech2018A&C....25...52C}.

\section{RFIm: a RFI Mitigation Library}\label{sec:software}

Our proposed solution to the challenge of real-time RFI mitigation is RFIm, a high-performance, open-source, tuned and extensible RFI mitigation library.
The goal of this library is to provide highly-tuned RFI mitigation routines that are designed to run in real-time on many-core accelerators, such as the many GPUs installed on ARTS, while at the same time be accurate enough to reduce false positives and increase the measured SNR of real astrophysical events.
Although RFIm is designed to be the main ARTS RFI mitigation library, it was also designed to be used on its own outside of the ARTS pipeline.

All mitigation methods are implemented in C++ and OpenCL, the Open Computing Language~\cite{munshi2009opencl}.
The C++ implementations are reference implementations, optimized for testing and readability, and are not used in production, while the OpenCL implementations are meant to be used in production, and are therefore highly optimized to run in real-time on GPUs.
The library is open source, released under version 2.0 of the Apache licence, and can be downloaded from GitHub\footnote{Available at https://github.com/AA-ALERT/RFIm}.

One of the main characteristics of RFIm is that it is designed with both code and performance portability in mind, and this is achieved using code generation and auto-tuning.
This means not only that the same code can be executed, without changes, on different hardware platforms, but also that the expected efficiency should be similar on these platforms.
In fact, the OpenCL code of the computational kernels is generated at run-time from a set of templates, so that it can be automatically tailored to the specific use-cases of the user, and it can take advantage of run-time knowledge of system and input parameters.

Moreover, a variety of parameters, such as the number of threads, the amount of work per thread, or synchronization schemes, can be tuned during code generation, so that the resulting code can achieve high performance running on different hardware.
This also allows processing efficiently data generated by different telescopes, or simply when observational parameters are changed because of different observations.
To help the user define all these tunable parameters, RFIm includes a tuning routine for each of the mitigation algorithms included.
By using these auto-tuning routines, a user can identify the best code generation parameters to use for a specific observational setup and computing platform, and save the resulting configurations for use in production.

RFIm currently contains two mitigation methods, both of them real-time thresholding methods: time-domain sigma cut (TDSC), and frequency-domain sigma cut (FDSC).
Further methods are being developed and will be integrated into RFIm in the future, including the Sum-threshold~\cite{offringa2012sumthreshold} and Edge-threshold~\cite{boyle2019} algorithms.

\section{Time and Frequency Domain Sigma Cut}\label{sec:tfdsc}

During observations, AMBER splits the input time-series in order to process data in batches, where the number of time samples inside a batch varies depending on the number of samples per second and the DM range being searched.
Therefore, RFI mitigation within RFIm is performed on a per-batch basis.
Below, we briefly describe the two algorithms currently implemented within RFIm. 

\begin{figure}
    \centering
    \includegraphics[height=0.35\textwidth]{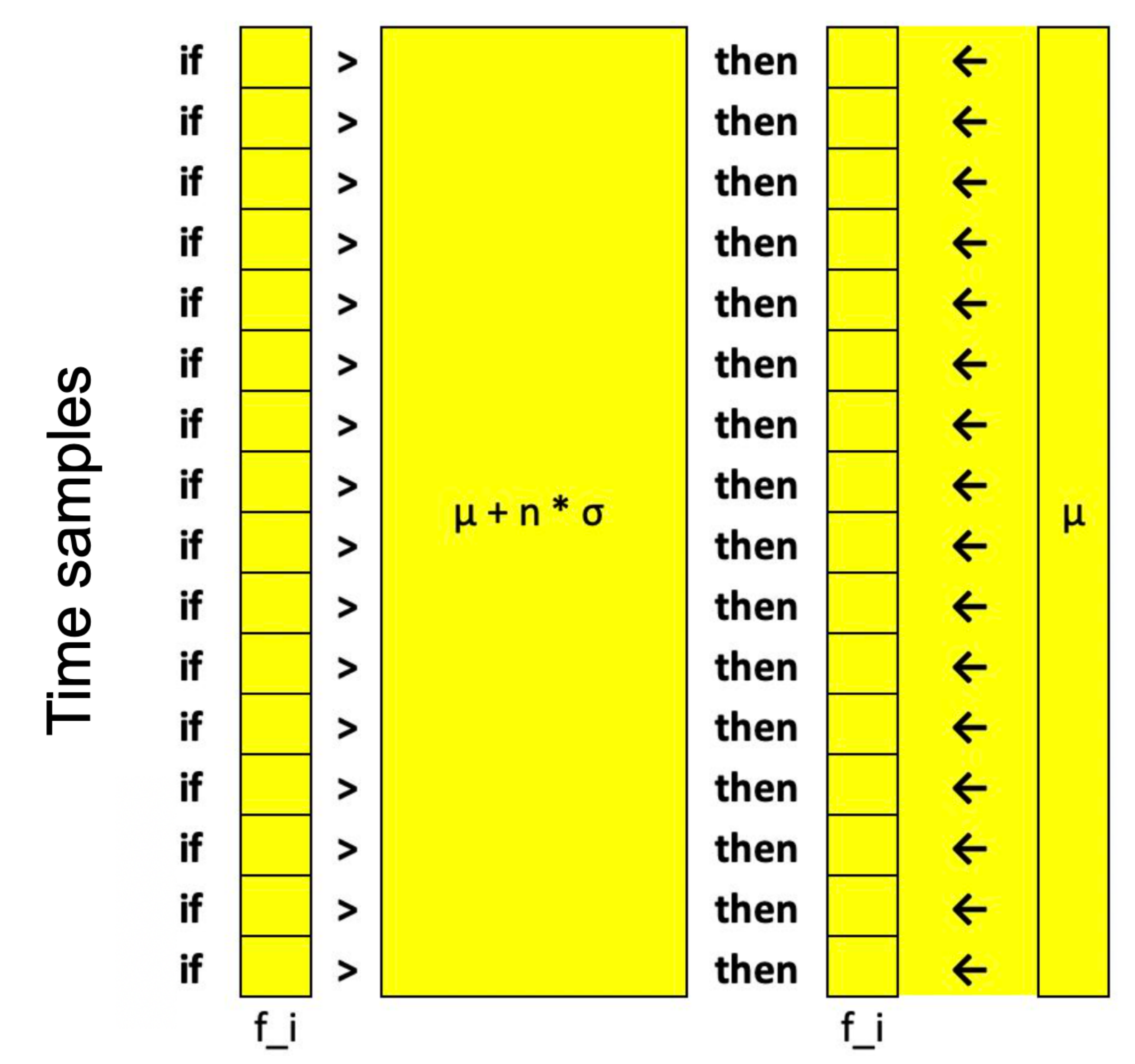}
    \caption{\textbf{Schematic diagram of the TDSC algorithm computed within an OpenCL kernel.} For a given frequency channel (f\_i), each time sample (t\_i) of a batch is compared to the statistics of the batch: mean ($\mu$), and $n$ times the standard deviation ($\sigma$). If a sample is an outlier within the batch, it is replaced by the mean of the batch. We note that in practice, the applied comparison is symmetric.}
    \label{fig::tdsc}
\end{figure}

The TDSC method (Figure~\ref{fig::tdsc}) targets bright broadband events of low-DM. For a given frequency channel (f\_i), each time sample (t\_i) of a batch is compared to the global statistics of the batch: the mean ($\mu$), and the standard deviation ($\sigma$). If t\_i is an outlier within the batch (e.g. $|\textrm{t\_i}-\mu| > n\sigma$, where $n$ is the user-defined threshold) it is replaced by the mean of the batch.

\begin{figure}
    \centering
    \includegraphics[height=0.35\textwidth]{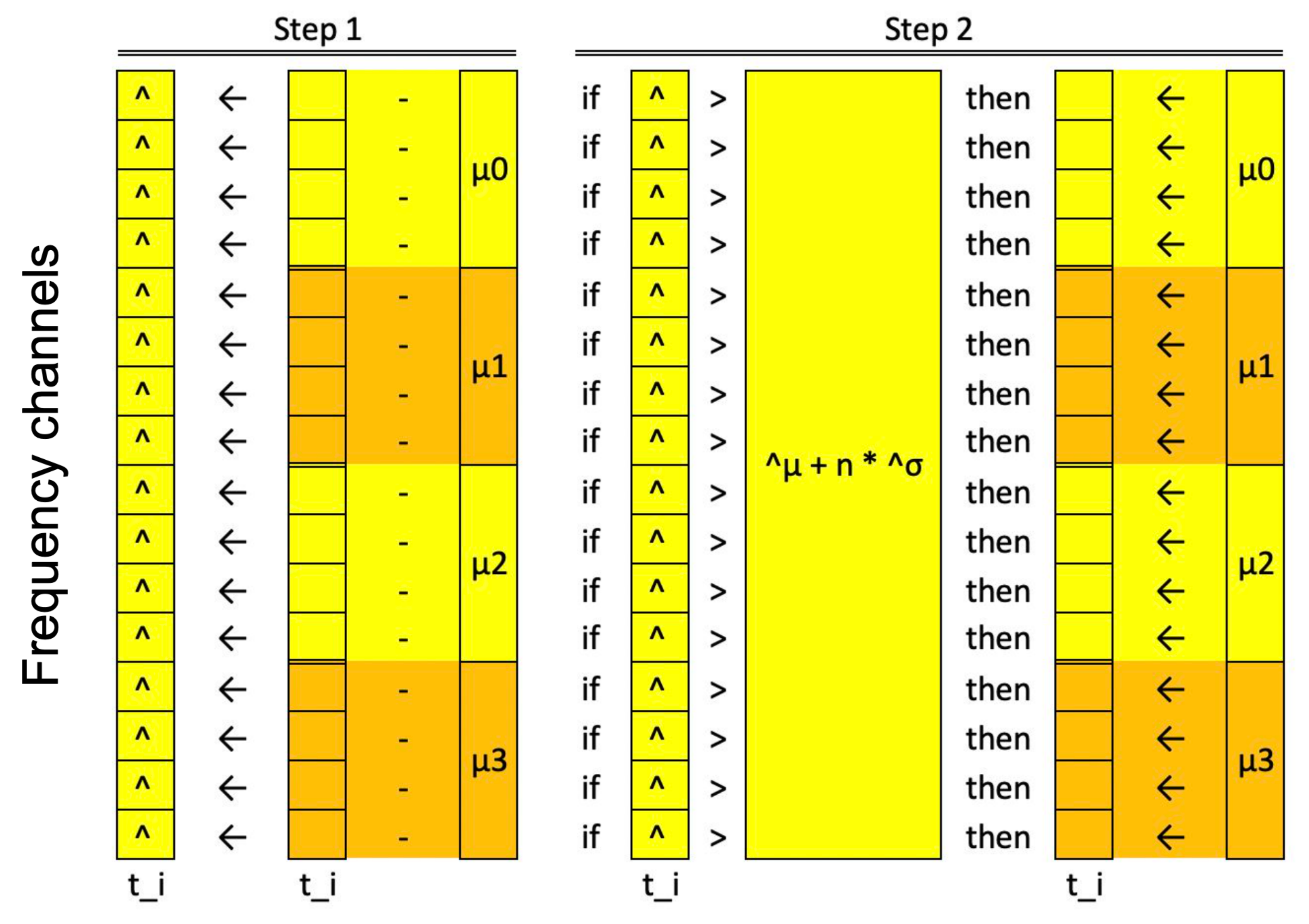}
    \caption{\textbf{Schematic diagram of the FDSC algorithm computed within an OpenCL kernel.} The algorithm proceeds in two steps: (1) for a given time sample (t\_i), each frequency channel (f\_i) is first attributed a bin, and a difference between each frequency channel and the mean of the bin is computed; (2) each mean-corrected frequency channel is compared to the mean-corrected statistics of all frequency channels following a similar process as in TDSC. If a mean-corrected sample is an outlier, the original channel value is replaced by the mean of its respective bin. We note that in practice, the applied comparison is symmetric.}
    \label{fig::fdsc}
\end{figure}

The FDSC method (Figure~\ref{fig::fdsc}) targets bright narrow-band pulses of short duration. FDSC comprises two steps. Firstly, for cases where no bandpass correction was applied to the data, frequency channels at a given time (t\_i) are grouped into $n$ bins, and the mean of the bin is subtracted from t\_i, resulting in a mean-corrected sample. Secondly, each mean-corrected frequency channel is compared to the mean-corrected statistics of all frequency channels following a similar process as described for TDSC. If a mean-corrected sample is an outlier, the original channel value is replaced by the mean of its respective bin.

Although the currently implemented replacement strategy for both TDSC and FDSC is to use the mean, this is done purely to reduce the computational cost of these algorithms.
In fact, both TDSC and FDSC support a parameter representing the replacement strategy for the flagged samples, and RFIm users can implement their own strategies, such as using the median, random unflagged samples, or any other user-defined strategy.

\section{Experimental Results}\label{sec:results}

\subsection{Accuracy}\label{sec:results:sub:accuracy}

\begin{figure}
    \centering
    \includegraphics[width=0.48\textwidth]{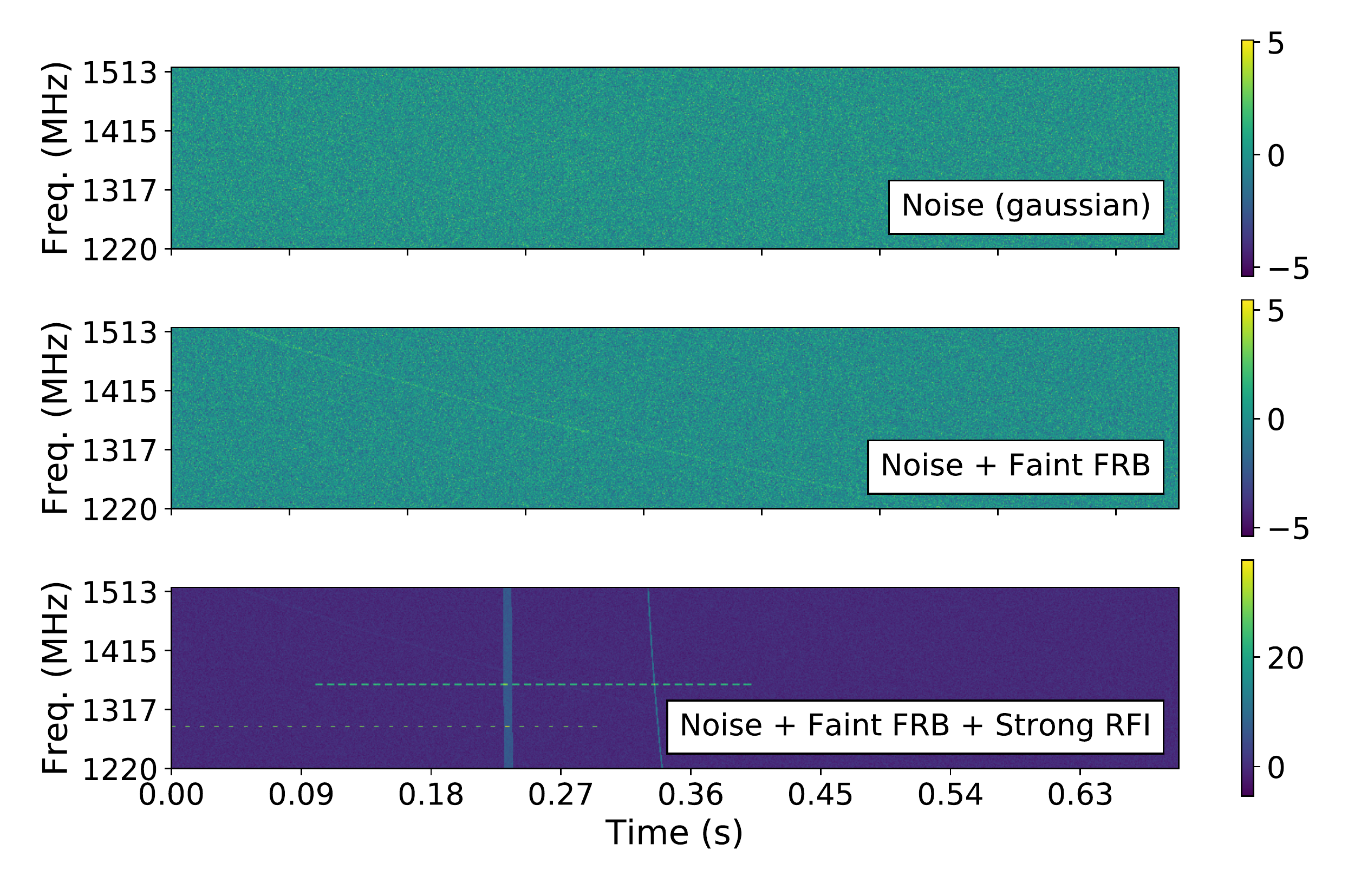}
    \caption{\textbf{Time-frequency plots of simulated data in the Apertif frequency band.} Top: gaussian noise; middle: one faint high DM pulse, with DM = 500~pc/cm$^3$ and SNR = 15 is added to the noise; bottom: two low DM pulses with $\textrm{DM}\in\{1, 10\}$ and $\textrm{S/N}\in\{100, 150\}$ respectively, and two periodic narrow band pulses with SNR = 125 covering 10, 15 frequency channels respectively. Data intensity (as highlighted in the colorbars) is in an arbitrary unit.}
    \label{fig::input}
\end{figure}

To evaluate how the methods proposed in the previous section perform we test the algorithms on both artificial and real data.
We begin with simulated data based on surveying parameters of ARTS (time sampling, total bandwidth, channelization, etc.). Using this canvas, we generate Gaussian noise data with $\mu=0$ and $\sigma=1$. We then add to the noise a faint, broadband highly dispersed pulse representing an FRB. Finally, we add strong RFI signals (two bright low-DM broadband pulses, and two narrow-band periodic pulses). Figure~\ref{fig::input} shows the test data generation procedure.

\begin{figure}
    \centering
    \includegraphics[width=0.48\textwidth]{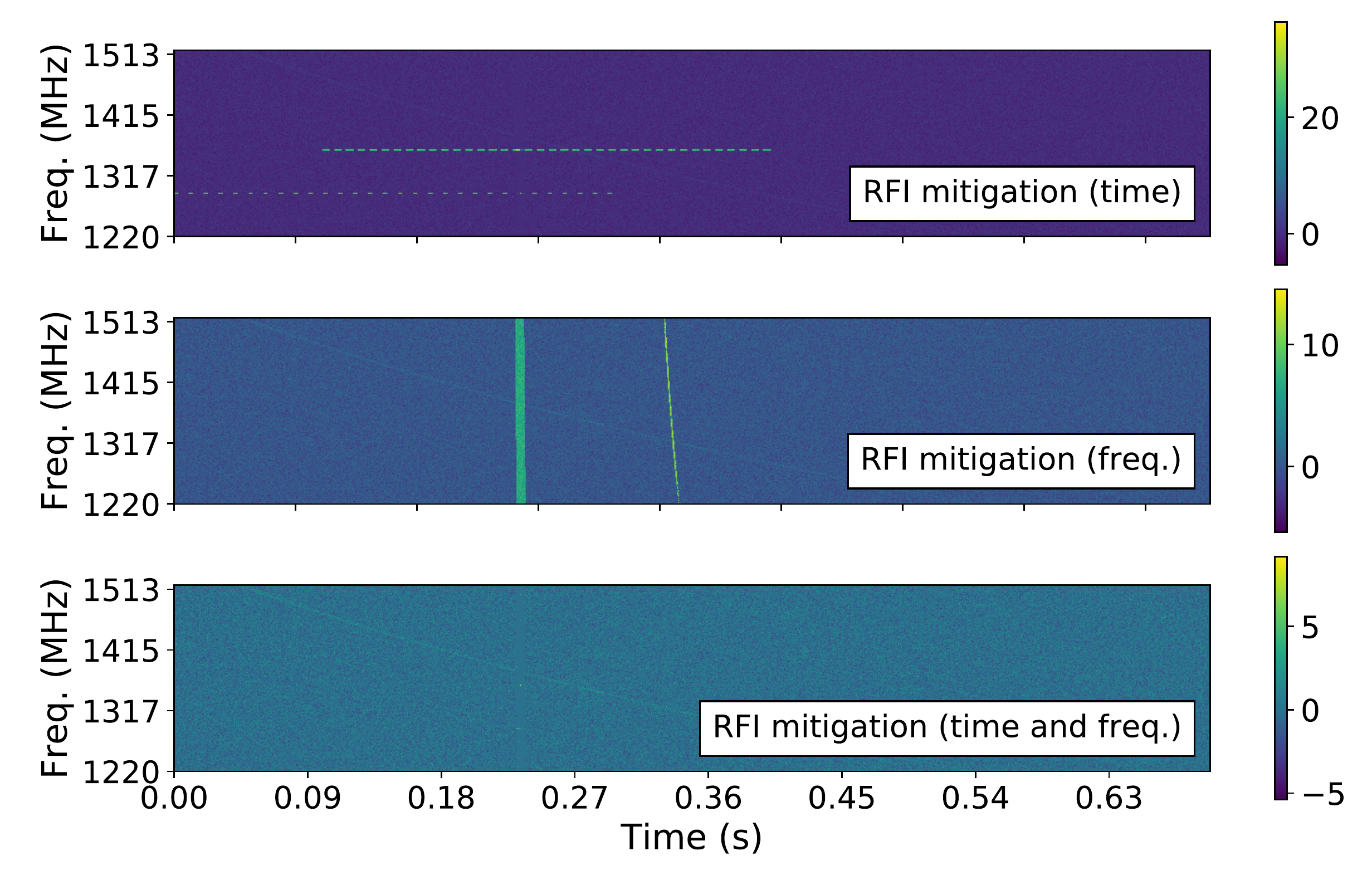}
    \caption{\textbf{RFI mitigation applied to data from the bottom panel of Fig. \ref{fig::input}.} From top to bottom: mitigation in time with threshold=3.25$\sigma$; mitigation in frequency with a bin size of 32 channels, and threshold=2.75$\sigma$; both mitigation methods applied (zero-DM plane); and dedispersed window (DM = 500~pc/cm$^3$) with both mitigation methods applied. The faint dispersed pulse can be seen starting from t=0.04s at the highest frequency.}
    \label{fig::examples}
\end{figure}

To highlight the behaviour of each mitigation technique, we proceed in applying our mitigation algorithms on the dispersed data (DM=0, Figure~\ref{fig::examples}): TDSC only (top panel), FDSC only (central panel), and both methods combined (bottom panel). As expected, bright broadband RFI is cleaned by the TDSC, and bright narrow-band RFI is cleaned by FDSC. When applying both techniques on the test data, the faint FRB becomes apparent. 

\begin{figure}
    \centering
    \includegraphics[width=0.48\textwidth]{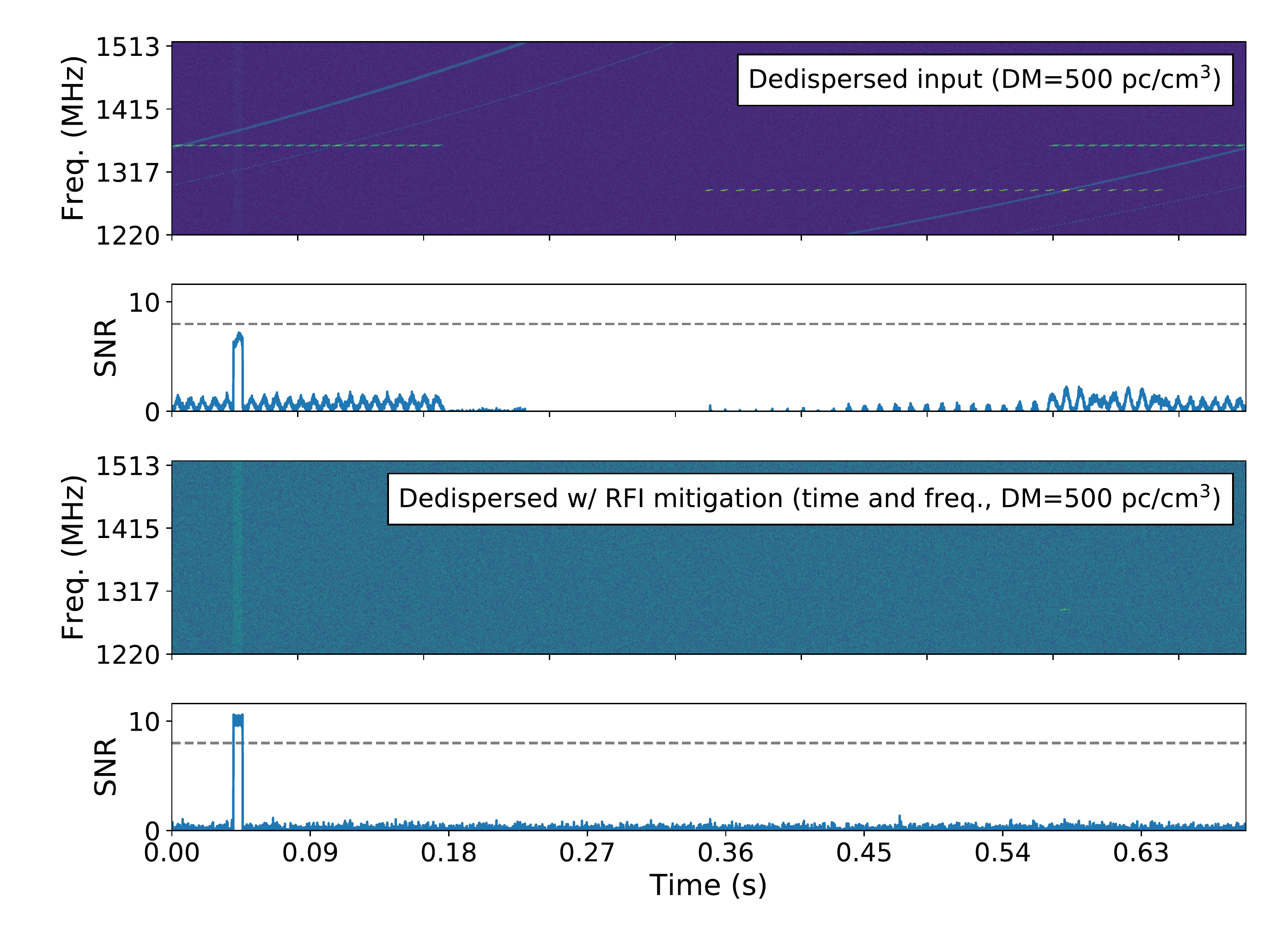}
    \caption{\textbf{Time series from Fig.~\ref{fig::input} with and without RFI mitigation dedispersed at FRB's DM.} The dashed line shows the threshold separating a trigger (signal going above the threshold) from a non-trigger. After RFI mitigation, this specific FRB is detected.}
    \label{fig::dedispersed}
\end{figure}

We then de-disperse the data to the FRB's DM value, and compare the result with and without RFI mitigation (Figure~\ref{fig::dedispersed}). In this specific case, the faint FRB would not have been detected by the pipeline as the SNR of the dedispersed burst is below the detection threshold (top two panels). On the other hand, the time-series cleaned with RFIm shows the dedispersed burst having a SNR greater than the detection threshold. 

Finally, to assure we do not radically remove high-DM pulses from observations, we evaluate our method on real data. We apply TDSC and FDSC to a 299 seconds Apertif observation of PSR B0531+21\footnote{PSR B0531+21 is commonly known as the Crab pulsar. The data is sampled at 81.92~$\mu$s, with 300~MHz of bandwidth at L-Band, and a channel width of ~0.195~MHz. It is common to see variations in SNR between individual pulses.} using the OpenCL implementation of AMBER. To simulate the iterative mitigation process used in the production pipeline, we perform three iterations of the algorithms. 

\begin{figure}
    \centering
    \includegraphics[width=0.48\textwidth]{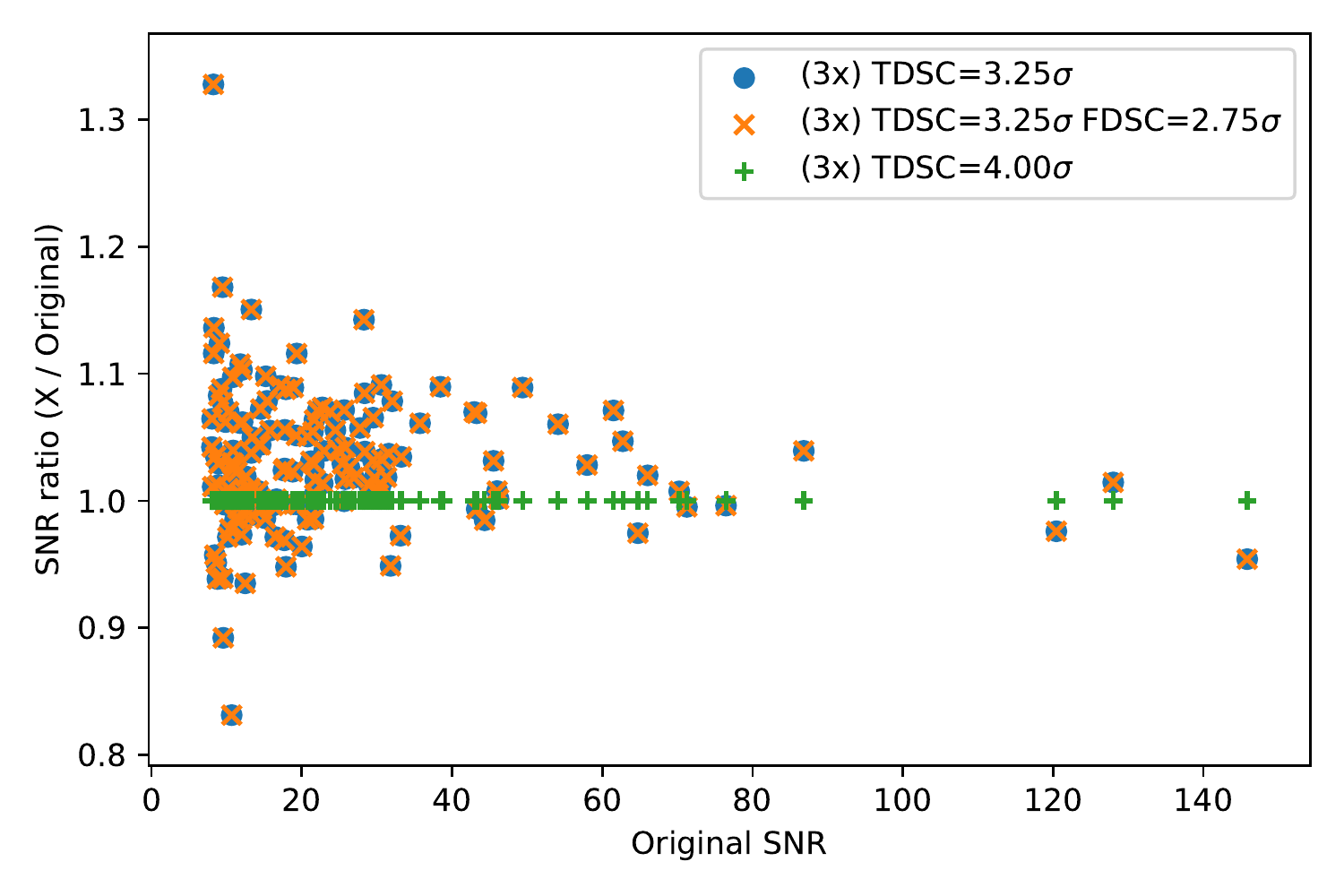}
    \caption{\textbf{Effect of applying TDSC and FDSC to detection's signal-to-noise ratio (SNR) with AMBER to an Apertif observation of PSR B0531+21.} The plot compares the SNR ratio as a function of the SNR computed using the original data (without RFI mitigation). The ratio is computed between SNR detected with and without RFI mitigation (X/Original), where X corresponds to one of three experimental cases, where we apply RFI mitigation using threshold values of: TDSC=3.25$\sigma$, combined TDSC=3.25$\sigma$ and FDSC=2.75$\sigma$, and TDSC=4.00$\sigma$ respectively.}
    \label{fig::crab}
\end{figure}

Figure~\ref{fig::crab} highlights the effect of mitigation on SNR for three experimental cases, where we apply RFI mitigation using threshold values of: TDSC=3.25$\sigma$, combined TDSC=3.25$\sigma$ and FDSC=2.75$\sigma$, and TDSC=4.00$\sigma$ respectively. The scatter plot shows the SNR ratio---the SNR computed after applying RFI mitigation divided by the SNR computed on the original data---for multiple single pulse detections, as a function of the original SNR (no mitigation applied). A ratio of 1 means there was no difference after applying mitigation for a given set of parameters; a ratio greater than 1 means an improved SNR after applying mitigation; and a ratio smaller than 1 means a reduced SNR after applying mitigation. 

It is interesting to note that applying TDSC with 3.25$\sigma$ (blue circles) improves the SNR in most cases (74\%), and that the largest differences are found close to our detection threshold of 8. Similarly, applying both TDSC with 3.25$\sigma$ and FDSC with 2.75$\sigma$ (orange x symbols) gives the same results as when using only TDSC with 3.25$\sigma$ (the blue circles and orange x symbols overlap)---which is expected as FDSC should not affect broadband signal. Unsurprisingly, applying TDSC with a poorly constraining 4.00$\sigma$ threshold (green + symbols) gives identical results as in the case where no RFI mitigation is being applied. 

We keep further analysis of trigger reduction factors, causes of specific SNR improvements or reductions, and the effect on true- and false-positive detections (e.g. precision and recall) as future work. However, we note that the total trigger counts prior to the machine learning classification step vary between the various cases: 200 triggers without RFI mitigation, 195 when applying TDSC=3.25$\sigma$, and 93 with the combination of TDSC=3.25$\sigma$ and FDSC=2.75$\sigma$---hence reducing the amount of data to be post-processed. Finally, we note some events detected in the original data are not detected after RFI mitigation with 3.25$\sigma$ (10 missed detections), and vice-versa (14 new detections previously undetected in the original data)---which is in agreement with our simulation results.

\subsection{Performance}\label{sec:results:sub:performance}

\begin{figure}
    \centering
    \includegraphics[width=0.48 \textwidth]{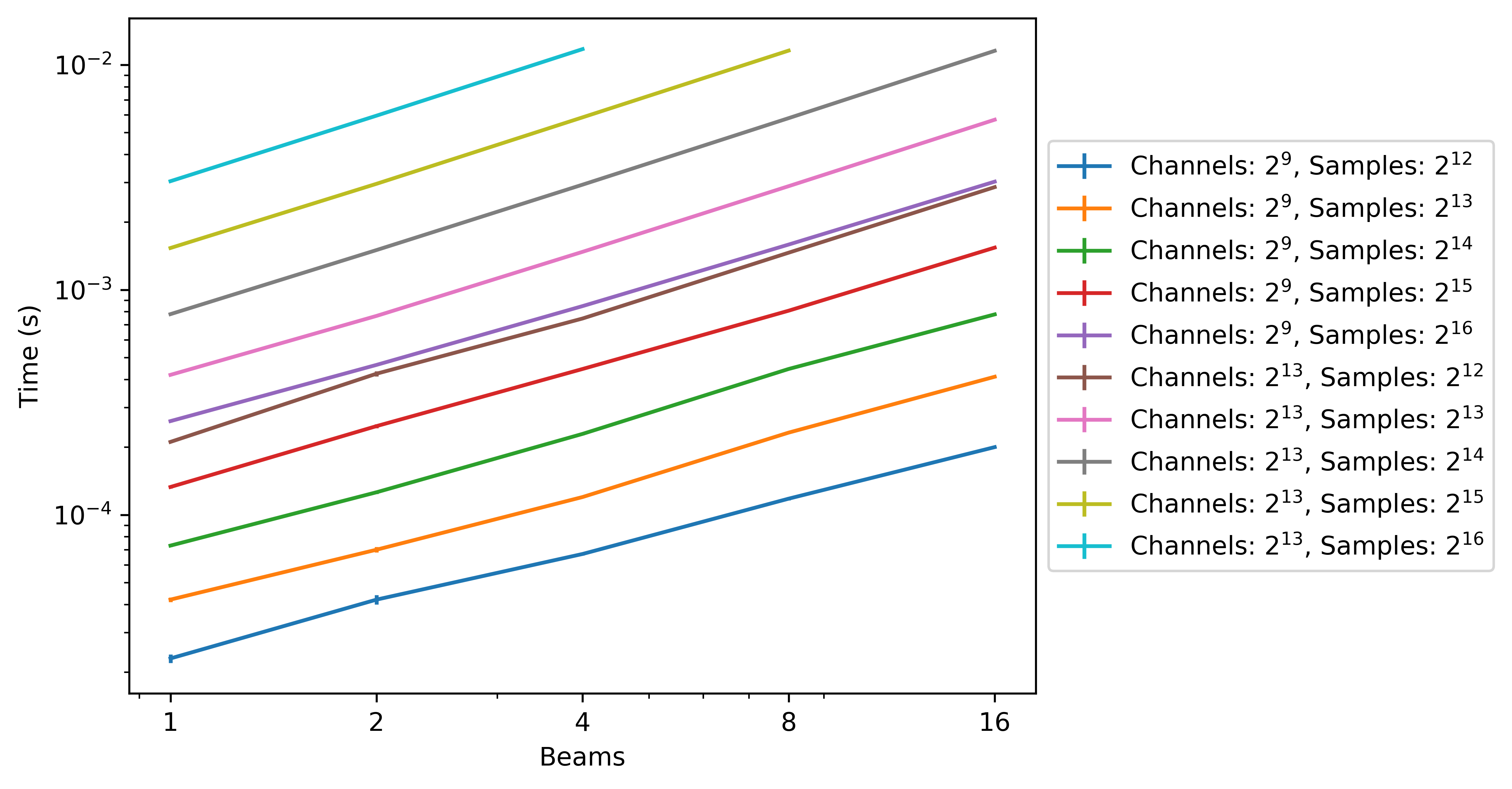}
    \caption{\textbf{Scalability experiment for the TDSC algorithm.}}
    \label{fig:tdsc_scalability}
\end{figure}

\begin{figure}
    \centering
    \includegraphics[width=0.48 \textwidth]{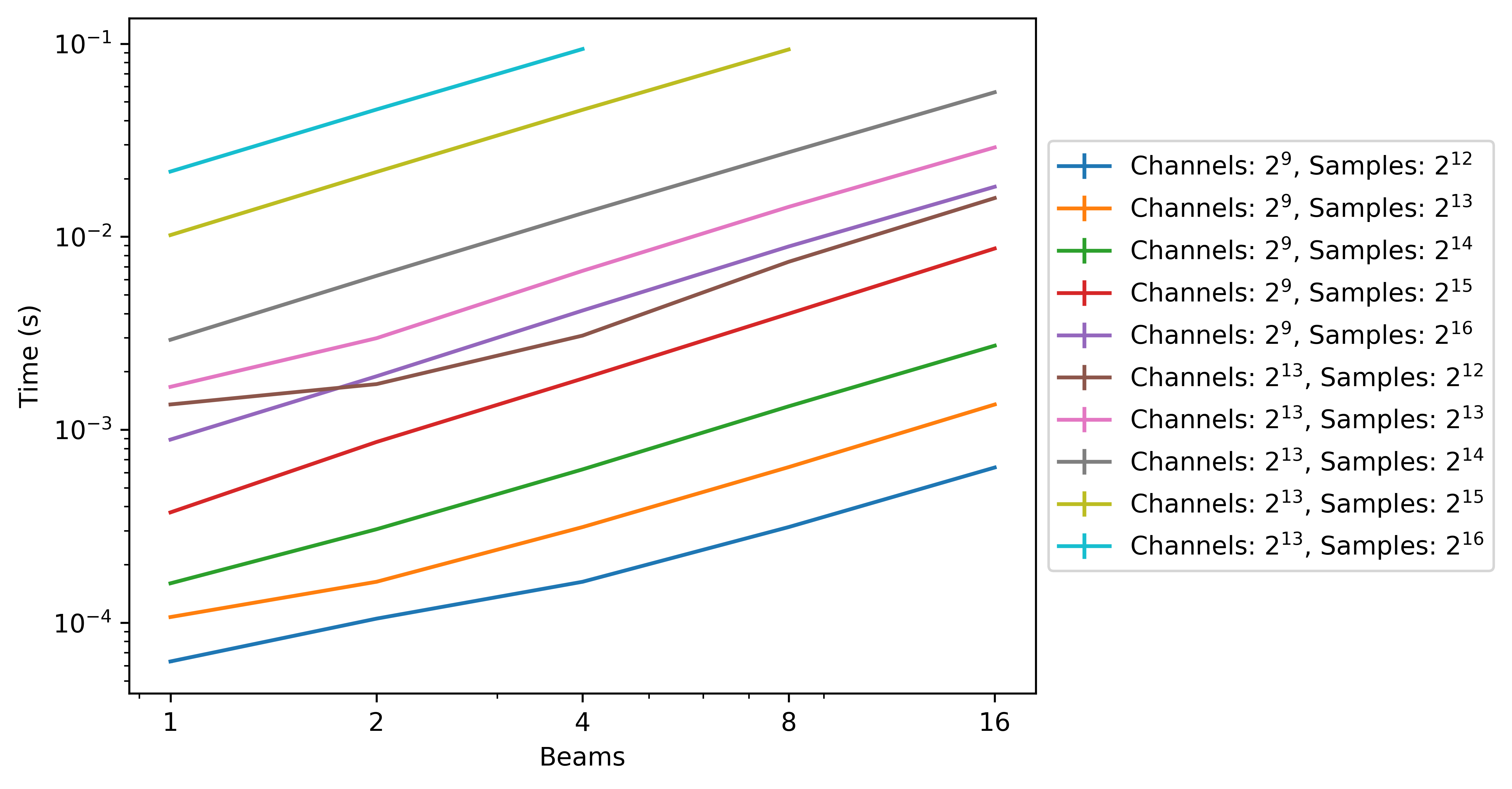}
    \caption{\textbf{Scalability experiment for the FDSC algorithm.}}
    \label{fig:fdsc_scalability}
\end{figure}

One of the design goals of RFIm is to provide accurate results in real-time, and while accuracy has been already discussed in Section~\ref{sec:results:sub:accuracy}, this section focuses on performance.
Figure~\ref{fig:tdsc_scalability} presents the scalability of the TDSC algorithm, while Figure~\ref{fig:fdsc_scalability} shows the same experiment for the FDSC algorithm.

In these experiments, tuned OpenCL implementations of the two algorithms are run on one of the NVIDIA GTX~1080Ti cards of the ARTS cluster, and execution time is measured; presented results are the average of multiple executions, to remove statistical fluctuations from the data.
The parameters that are varied in the experiment are: (1) the number of frequency channels, (2) the number of time samples, and (3) the number of input beams; to improve the readability of the figures, a representative selection of the results is plotted.
In both Figure~\ref{fig:tdsc_scalability} and~\ref{fig:fdsc_scalability}, each color coded line represents the execution time for a given combination of number of frequency channels and time samples, in correspondence with the number of beams on the $x$ axis; both the $x$ and the $y$ axes are in logarithmic scale.

The first noticeable result is that both algorithms run in real-time, with the execution time being, in the most complex cases, at most a tenth of a second for a whole batch of input data.
This is important because, in the context of an FRB searching pipeline, most of the processing time can still be dedicated to the search itself.
Another result is that both algorithms scale linearly not just in terms of number of beams, but also in terms of number of frequency channels and time samples.
Therefore, an increase in sensitivity or resolution in WSRT, or any other telescope for which RFIm is used, can more easily be accommodated, without requiring major changes in the software.

\begin{figure}
    \centering
    \includegraphics[width=0.48 \textwidth]{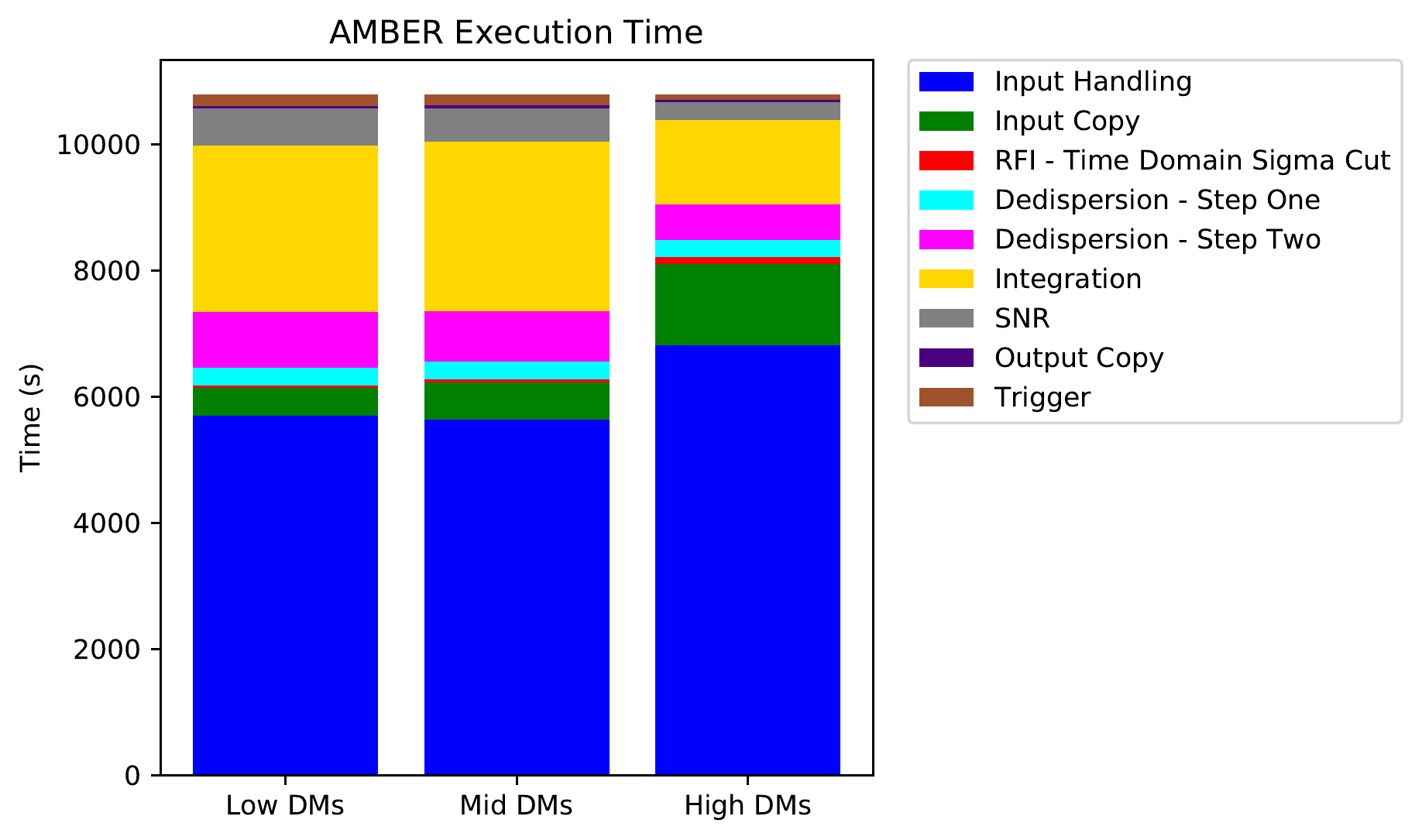}
    \caption{\textbf{Median execution time of three different instances of AMBER.} RFI mitigation (in red) takes between 0.3\% and 1\% of the total execution time.}
    \label{fig:amber_execution_time}
\end{figure}

Another way to look at the performance of RFIm is to measure its contribution to the total execution time of AMBER, the ARTS FRB pipeline. The timing data was recorded during a survey observation in July 2019.
Figure~\ref{fig:amber_execution_time} presents an overview of the time, in seconds, spent inside each component of the AMBER pipeline, for this three hours observation; the three columns in the figure correspond to three different instances of AMBER, each running on a subset of the DM space.

During this observation, the TDSC algorithm of RFIm was enabled, and three iterations with 3.25$\sigma$ were executed for each batch; the number of frequency channels is 1536, the number of beams 12, and the number of samples per second 12,208.
Results show that the contribution of TDSC to AMBER's execution time varies between the 0.33\% for the low DMs search, to the 1.02\% of the high DMs search.
Therefore, using RFIm we can mitigate the effects of RFI for ARTS, without significantly affecting the execution time of the whole pipeline.

\section{Conclusions}\label{sec:conclusions}

RFI will be an important problem for all upcoming FRB surveys, including the Square Kilometre Array. As such, there is a need for real-time RFI mitigation.
In this paper we introduced RFIm, a high-performance, open-source, tuned and extensible RFI mitigation library, and showed how it can improve the quality of results with real and simulated data.
Moreover, we showed that RFIm achieves real-time performance in different scenarios (varying number of frequency channels, number of time samples per batch, number of input beams). RFIm is already used in production for ARTS, mitigating RFI while barely affecting the execution time of AMBER.

While we aim to mitigate RFI, there exists a trade-off between pre-dedispersion RFI cleaning and false-positive rejection post-triggering by the deep-learning classifier. To all current and future actions to remove RFI during real-time pulsar and FRB searches, the risk of over-cleaning the data and removing astronomical events needs to be considered~\cite{Connor+2018AJ....156..256C}.
With this in mind, we note that RFIm mitigation actions are currently performed in GPU memory only, leaving the original data unmodified, while the FRB search is performed on the mitigated data stream in memory. Hence, provided the raw data is stored to disk during observation, one can also process the data \emph{a posteriori} using different cleaning methods for further analysis.

\section*{Acknowledgments}

This work is supported by the Netherlands eScience Center under the project ``AA-ALERT: Access and Acceleration of the Apertif Legacy Exploration of the Radio Transient Sky'', filenumber 027.015.G09.
We would like to thank Liam Connor, Joeri van Leeuwen, and Yogesh Maan for valuable discussions during the development of RFIm.

\bibliographystyle{IEEEtran}
\bibliography{rfi2019_proc}

\end{document}